\documentclass[a4paper, 11pt]{article}
\usepackage[english]{babel}
\usepackage{setspace}
\usepackage{graphicx}
\usepackage{float}
\usepackage{enumerate}
\usepackage{url}
\usepackage{amsmath}
\usepackage[super,numbers,sort&compress]{natbib}
\usepackage{hyperref}
\usepackage{cleveref}
\usepackage[T1]{fontenc}
\usepackage{verbatim}
\usepackage{braket}
\usepackage[affil-it]{authblk}
\usepackage[textheight=23cm,textwidth=17cm]{geometry}
\usepackage{subcaption}

\author{Salom\'e R. Rieder} % ORC-ID: 0000-0001-8036-651X
\author{Benjamin Ries} % ORC-ID: 0000-0002-0945-8304 
\author{Kay Schaller} % ORC-ID: 0000-0003-1555-5626
\author{Candide Champion} % ORC-ID: 0000-0001-6982-1569
\author{Emilia P. Barros} % ORC-ID:  0000-0001-9755-260X 
\author{Philippe H. H\"unenberger\thanks{Co-corresponding author: email: phil@igc.phys.chem.ethz.ch, ORCID: 0000-0002-9420-7998}} % ORC-ID: 0000-0002-9420-7998
\author{Sereina Riniker\thanks{Co-corresponding author: email: sriniker@ethz.ch, ORCID: 0000-0003-1893-4031}} % ORC-ID: 0000-0003-1893-4031
\affil{Laboratory of Physical Chemistry, ETH Zurich, Vladimir-Prelog-Weg 2, 8093 Zurich, Switzerland}

\title{\textbf{RE-EDS Using GAFF Topologies: Application to Relative Hydration Free-Energy Calculations for Large Sets of Molecules}}

\date{}

\begin{document}

\maketitle

\begin{abstract}
Free-energy differences between pairs of end-states can be estimated based on molecular dynamics (MD) simulations using standard pathway-dependent methods such as thermodynamic integration (TI), free-energy perturbation, or Bennett's acceptance ratio. Replica-exchange enveloping distribution sampling (RE-EDS), on the other hand, allows for the sampling of multiple end-states in a single simulation without the specification of any pathways. In this work, we use the RE-EDS method as implemented in GROMOS together with generalized AMBER force field (GAFF) topologies, converted to a GROMOS-compatible format with a newly developed GROMOS++ program \textit{amber2gromos}, to compute relative hydration free energies for a series of benzene derivatives. The results obtained with RE-EDS are compared to the experimental data as well as calculated values from the literature. In addition, the estimated free-energy differences in water and in vacuum are compared to values from TI calculations carried out with GROMACS. The hydration free energies obtained using RE-EDS for multiple molecules are found to be in good agreement with both the experimental data and the results calculated using other free-energy methods. While all considered free-energy methods delivered accurate results, the RE-EDS calculations required the least amount of total simulation time. This work serves as a validation for the use of GAFF topologies with the GROMOS simulation package and the RE-EDS approach. Furthermore, the performance of RE-EDS for a large set of 28 end-states is assessed with promising results.
\end{abstract}

\section{Introduction}
In recent years, free-energy calculations (either absolute or relative) based on classical molecular dynamics (MD) simulations have started to play an increasingly important role in the field of computer-aided drug design.\cite{Michel2010,Borhani2012,deVivo2016,Cournia2017,WilliamsNoonan2018,Armacost2020,Cournia2020,Lee2020,Song2020} There exist many well-established pairwise free-energy methods such as
thermodynamic integration (TI),\cite{Kirkwood1935}
free-energy perturbation (FEP),\cite{Zwanzig1954}
Bennett's acceptance ratio (BAR),\cite{Bennett1976} and
multistate BAR (MBAR).\cite{Shirts2008} Approaches such as multi-site $\lambda$-dynamics\cite{Knight2011,Raman2020,Hayes2021} and enveloping distribution sampling (EDS)\cite{Christ2007,Christ2008} enable the calculation of pairwise free-energy differences for multiple end-states from a single simulation. While $\lambda$-dynamics uses the coupling parameter $\lambda$ as a dynamic variable to connect the end-states, EDS is a pathway-independent method that samples a reference state ``enveloping'' all end-states.
Recently, replica-exchange EDS (RE-EDS)\cite{Sidler2016,Sidler2017,Ries2021} and accelerated EDS (A-EDS)\cite{Perthold2018,Perthold2020} have been developed as extensions of EDS to simplify the parameter optimization and improve the performance of EDS. Both methods are implemented in the GROMOS software package.\cite{Schmid2012}

%topologies and MD software file formats
Apart from the free-energy method, the quality of the underlying force field is crucial for the accuracy of free-energy calculations, and of MD simulations in general.\cite{Hunenberger1999,Nerenberg2018,Monticelli2013,Riniker2018} In the past years, various tools have been developed to automate the otherwise laborious task of topology generation for small molecule ligands, such as 
antechamber,\cite{Wang2001,Wang2004,Wang2006} the automated topology builder (ATB),\cite{Mark2011,Stroet2018} the fragment-based CombiFF,\cite{Oliveira2020,Oliveira2021} general automated atomic model parameterization (GAAMP),\cite{Huang2013,Boulanger2018} LigParGen,\cite{Dodda2017} open force field (SMIRNOFF and OpenFF),\cite{Mobley2018,Qiu2021} ParamChem,\cite{Vanommeslaeghe2010,Vanommeslaeghe2012,Vanommeslaeghe2012a} PRODRG,\cite{Schuttelkopf2004} R.E.D.,\cite{Vanquelef2011} or SwissParam.\cite{Zoete2011}
MD simulation engines such as
AMBER,\cite{Weiner1981,Salomon2013,AMBER_manual_2016}
CHARMM,\cite{Brooks1983,Brooks2009}
GROMACS,\cite{Bekker1993,Abraham2015}
GROMOS,\cite{gromos96,Schmid2012}
or OpenMM\cite{Eastman2010,Eastman2017} require specific file formats to describe the system topology and coordinates. In addition, there are also small differences in the functional form of the force fields or in the units used by different MD engines.\cite{Riniker2018} In many cases, tools are already available to translate between some of the different file formats, enabling e.g. the use of AMBER topologies with GROMACS.\cite{Swails2010,daSilva2012,gromos2amber,Vermaas2016,Hedges2019}

%motivation for hydration free energies
The calculation of (absolute or relative) hydration free energies serves as a straightforward test case to assess and compare the quality of different free-energy methods and to validate force fields.\cite{Guthrie2009,Kashefolgheta2020} Thanks to databases such as FreeSolv,\cite{Mobley2014,Duarte2017} the Minnesota solvation database,\cite{Marenich2012} or the ATB server,\cite{Mark2011,Stroet2018} ample reference data is available for both experimental results as well as calculated values obtained with different force fields and free-energy methods. Furthermore, the calculation of hydration free energies is computationally far less expensive than, for example, that of binding free energies.

%%outlook
In this work, a newly introduced GROMOS++\cite{Eichenberger2011} program \textit{amber2gromos}, developed by the authors of this study, is described. It translates a topology from the AMBER prmtop\cite{Swails2013} file format to a GROMOS topology, enabling users of the GROMOS MD engine to simulate systems with the AMBER or generalized AMBER (GAFF\cite{Wang2004}) force fields. Extension to the OpenFF\cite{Mobley2018,Qiu2021} family of force fields is straightforward. In the following, the underlying differences between the AMBER and GROMOS force fields are discussed, and the necessary conversions are described in detail. The correctness of the topology conversion is validated by comparison of single-molecule simulations in vacuum using GROMACS or GROMOS. Furthermore, two sets of small benzene derivatives are assembled from the FreeSolv\cite{Mobley2014,Duarte2017} database: a small set of 6 molecules, labeled A, and a larger set of 28 molecules, labeled B. For these molecules, relative hydration free energies are calculated with TI\cite{Kirkwood1935} in GROMACS (set A only) and with RE-EDS\cite{Sidler2016,Sidler2017,Ries2021} in GROMOS (sets A and B). The results are compared to each other and to the experimental and calculated values reported in the FreeSolv\cite{Mobley2014,Duarte2017} database.

%%%%%%%%%%%% THEORY %%%%%%%%%%%%%%
\section{Theory}

\subsection{Differences between the AMBER and GROMOS Force Fields}

%% overview 
The use of an automation tool such as AmberTools\cite{AMBER_manual_2016} (i.e., \textit{antechamber}\cite{Wang2001,Wang2004,Wang2006} and \textit{tleap}) simplifies the process of topology generation for small organic molecules considerably. In order to use GAFF\cite{Wang2004} topologies in GROMOS, they have to be converted to a GROMOS-compatible file format. There are several differences between the AMBER/GAFF and GROMOS\cite{Oostenbrink2004,Schmid2011,Horta2016} force fields.

%% general differences
First, GAFF is an all-atom force field, whereas most GROMOS (compatible) force fields use united atoms (i.e., implicit hydrogens) for the aliphatic CH$_n$ groups, to reduce the computational cost.\cite{Riniker2018} The GROMOS force fields are usually parameterized with the simple point-charge (SPC)\cite{Berendsen1981} water model, whereas the AMBER force-field family is parameterized with the TIP3P\cite{Jorgensen1983} water model.\cite{Riniker2018} A minor difference is the use of different units, e.g. nm, degrees, and kJ~mol$^{-1}$ in GROMOS versus \AA, radians, and kcal~mol$^{-1}$ in AMBER.\cite{GROMOS_manual_v4, AMBER_manual_2016}

%% potential energy function
%%% covalent terms
Second, there are several differences in the potential-energy function, i.e. the functional form of the force fields. Here, the subscripts ``A'' (AMBER) and ``G'' (GROMOS) are used to distinguish the terms/parameters of the two force-field families. In AMBER, harmonic bond stretching and bond-angle bending terms are used,\cite{AMBER_manual_2016,Wang2004,Riniker2018}
\begin{align}
    V_{A,i}^\text{bond,harm}(d_i) &= K_{A,i}^{b,\text{harm}}\left(d_i - d_{0,i}\right)^2
    \label{eq: bond_amber}\\
    V_{A,i}^\text{angle,harm}(\theta_i) &= K_{A,i}^{a,\text{harm}}\left(\theta_i - \theta_{0,i}\right)^2 \, ,
    \label{eq: angle_amber}
\end{align}
where $d_i$ is the distance between two bonded atoms, $K_{A,i}^{b,\text{harm}}$ the harmonic bond force constant, $d_{0,i}$ the equilibrium distance, $\theta_i$ the angle formed by three bonded atoms, $K_{A,i}$ the harmonic angle force constant, and $\theta_{0,i}$ is the reference bond angle. In GROMOS, harmonic bond stretching and bond-angle bending are also implemented. However, quartic bond stretching and cosine-harmonic bond-angle bending are used by default to increase computational efficiency.\cite{Riniker2018} They are defined as follows\cite{GROMOS_manual_v2}
\begin{align}
    V_{G,i}^\text{bond,harm}(d_i) &= \frac{1}{2} K_{G,i}^{b,\text{harm}}\left(d_i - d_{0,i}\right)^2
    \label{eq: bond_gromos_harm}\\
    V_{G,i}^\text{angle,harm}(\theta_i) &= \frac{1}{2}K_{G,i}^{a, \text{harm}}\left(\theta_i - \theta_{0,i}\right)^2 
    \label{eq: angle_gromos_harm}\\
    V_{G,i}^\text{bond,quart}(d_i) &= \frac{1}{4}K_{G,i}^{b,\text{quart}}\left(d_i^2 - d_{0,i}^2\right)^2
    \label{eq: bond_gromos_quart}\\
    V_{G,i}^\text{angle,cos}(\theta_i) &= \frac{1}{2}K_{G,i}^{a, \text{cos}}\left(\cos(\theta_i)-\cos(\theta_{0,i})\right)^2 \, ,
    \label{eq: angle_gromos_cos}
\end{align}
with the parameters being defined analogously to the AMBER parameters. It is important to note that for AMBER/GAFF, the factor $1/2$ in the harmonic bond stretching and bond-angle bending equations is already included in the force constants $K_{A,i}^{b,\text{harm}}$ and $K_{A,i}^{a, \text{harm}}$ (compare Eqs. (\ref{eq: bond_amber}) and (\ref{eq: angle_amber}) with Eqs. (\ref{eq: bond_gromos_harm}) and (\ref{eq: angle_gromos_harm})).\cite{Wang2004,AMBER_manual_2016} The harmonic force constants can be converted to the quartic and cosine-harmonic force constants, respectively, as\cite{GROMOS_manual_v2}
\begin{align}
    K_{G,i}^{b,\text{quart}} &= \frac{K_{G,i}^{b,\text{harm}}}{2d_{0,i}^2}
    \label{eq: conversion_bond}\\
    \begin{split}
        K_{G,i}^{a,\text{cos}} &= \frac{2k_BT_\text{eff}}{\left[\cos\left(\theta_{0,i}+\left(\frac{k_BT_\text{eff}}{K_{G,i}^{a,\text{harm}}}\right)^{\frac{1}{2}}\right)-\cos \theta_{0,i}\right]^2}
        \\&+\frac{2k_BT_\text{eff}}{\left[\cos\left(\theta_{0,i}-\left(\frac{k_BT_\text{eff}}{K_{G,i}^{a,\text{harm}}}\right)^{\frac{1}{2}}\right)-\cos \theta_{0,i}\right]^2}
        \label{eq: conversion_angle}\, ,
    \end{split}
\end{align}
where $k_B$ is the Boltzmann constant and $T_\text{eff}$ is an effective absolute temperature for the conversion (e.g., 300~K).\cite{GROMOS_manual_v2}
Another difference between AMBER and GROMOS is the potential-energy function used for the out-of-plane distortions. In AMBER, the same function is used for both proper and improper dihedral changes,\cite{Wang2004,AMBER_manual_2016,Riniker2018}
\begin{align}
    V_{A,i}^\text{tors/imp}(\theta_i) &= K_{A,i}^\text{tors/imp}\left[1 + \cos(m\theta_i - \theta_{0,i})\right] \, .
    \label{eq: dihedral_amber}
\end{align}
In contrast, GROMOS uses different functional forms for proper and improper dihedral changes,\cite{Riniker2018,GROMOS_manual_v2}
\begin{align}
    V_{G,i}^\text{tors}(\theta_i) &= K_{G,i}^\text{tors}\left[1 + \cos(m\theta_i - \theta_{0,i})\right]\\
    V_{G,i}^\text{imp}(\xi_i) &= \frac{1}{2} K_{G,i}^\text{imp}\left(\xi_i - \xi_{0,i}\right)^2 \, .
    \label{eq: dihedral_gromos}
\end{align}
The improper term is also used for out-of-tetrahedron distortions around the CH$_1$ united atom.
%% non-bonded terms
Both force-field families use the Lennard-Jones functional form for the van der Waals interactions\cite{Wang2004,AMBER_manual_2016,GROMOS_manual_v2}
\begin{align}
    V_{A,ij}^\text{vdW} &= \left(\frac{A_{ij}}{r_{ij}^{12}} - \frac{B_{ij}}{r_{ij}^6}\right)
    \label{eq: vdw_amber}
    \\
    V_{G,ij}^\text{vdW} &= \left(\frac{C_{12,ij}}{r_{ij}^{12}}-\frac{C_{6,ij}}{r_{ij}^{6}}\right) \,,
    \label{eq: vwd_gromos}
\end{align}
where $r_{ij}$ is the (minimum-image) distance between atoms $i$ and $j$, $A_{ij}$ and $C_{12,ij}$ are the repulsion coefficients, and $B_{ij}$ and $C_{6,ij}$ are the dispersion coefficients. There are, however, differences in the combination rules used (geometric in GROMOS and Lorentz-Berthelot in AMBER\cite{Riniker2018}) and the handling of third-neighbor interactions. In AMBER/GAFF, the Lennard-Jones 1,4-interactions are scaled by a factor $1/2$,\cite{Wang2004,AMBER_manual_2016} whereas GROMOS force fields contain a special set of parameters\cite{GROMOS_manual_v6} ($CS_{12}$ and $CS_{6}$) for such interactions, typically involving a reduced repulsion coefficient. Using a scaling factor or reduced interaction parameters for third-neighbor interactions avoids having a too large repulsion in \textit{gauche} conformations (relative to trans conformations).\cite{GROMOS_manual_v2} In AMBER/GAFF, electrostatic 1,4-interactions are also scaled by a factor $1/1.2$.\cite{Wang2004,AMBER_manual_2016} Such a scaling is not applied in GROMOS, although in some cases, third neighbors are excluded completely, for example for atoms that are in or attached to an aromatic ring.\cite{GROMOS_manual_v6} Furthermore, in AMBER/GAFF, the factor\cite{Swails2013}
\begin{align}
    k_e^{1/2} = \left(\frac{1}{4\pi\varepsilon_0}\right)^{1/2} = 18.2223 \text{~[kcal~\AA~mol}^{-1}e^{-2}\text{]}^{1/2}
    \label{eq: coulomb_constant}
\end{align}
is included in the atomic charges for computational efficiency. Here, $k_e$ is Coulomb's constant and $\varepsilon_0$ is the permittivity of vacuum.

\subsection{Relative Hydration Free Energies}

\begin{figure}[h]
    \centering
    \includegraphics[width=\textwidth]{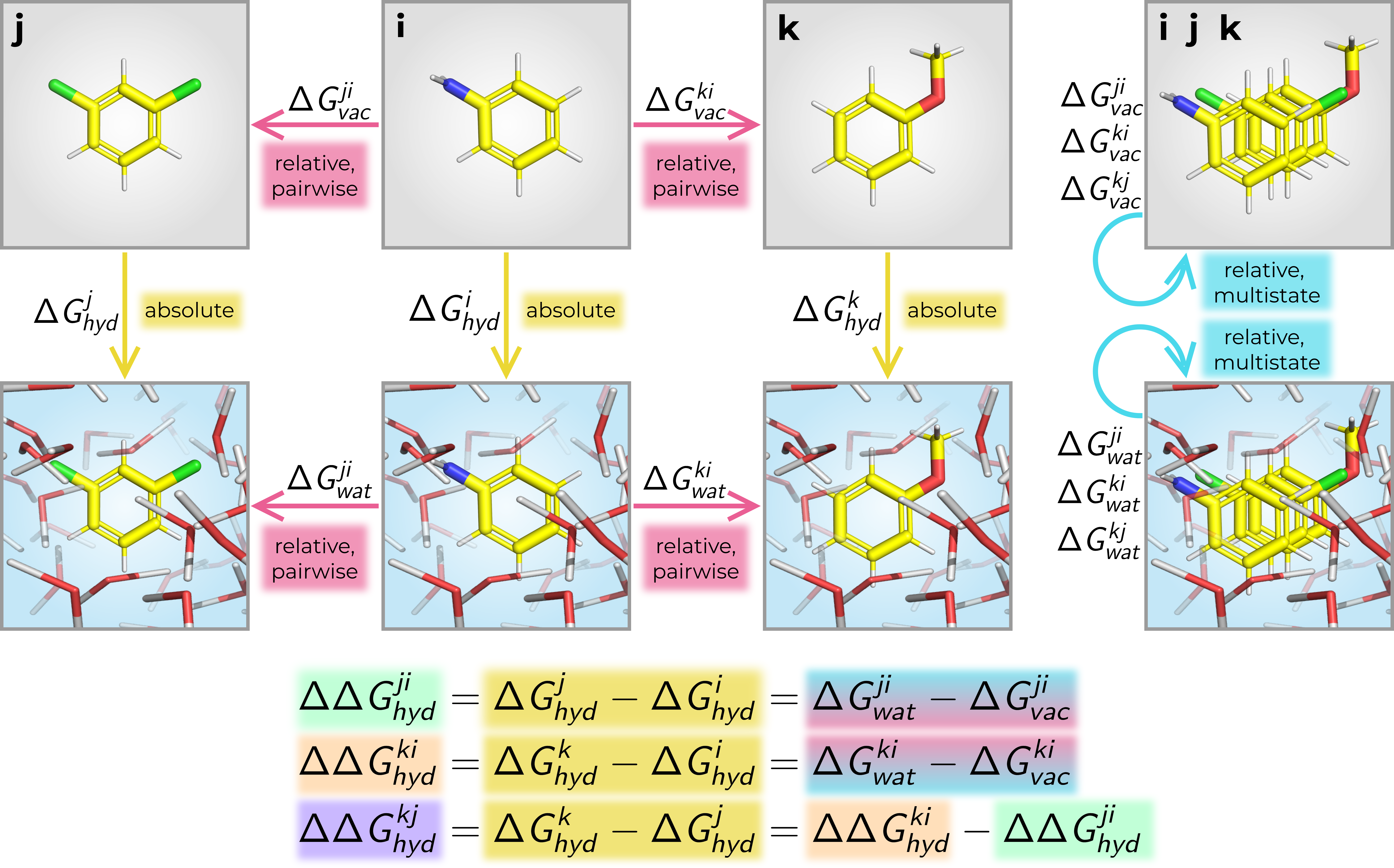}
    \caption{Thermodynamic cycle to calculate relative hydration free energies $\Delta\Delta G_\text{hyd}$ for three small molecules $i$ (aniline), $j$ (1,3-dichlorobenzene), and $k$ (anisole). Here, $\Delta G_\text{hyd}^i$ is the hydration free energy of molecule $i$, $\Delta G_\text{vac}^{ji}$ is the free-energy difference between molecules $i$ and $j$ in vacuum, $\Delta G_\text{wat}^{ji}$ is the free-energy difference between the two molecules in water, and $\Delta\Delta G_{hyd}^{ji}$ is the hydration free-energy difference between the two molecules (relative hydration free energy). The free-energy difference between two molecules can be calculated from multiple pairwise simulations (as shown on the left, e.g. with TI) or from one simulation with multiple molecules (as shown on the right, e.g. with (RE-)EDS).}
    \label{fig: thermodynamic_cycle}
\end{figure}

The hydration free energy quantifies the free-energy change when a molecule is transferred from gas to water.\cite{Guthrie2009,Klimkovich2010} In this work, relative hydration free energies $\Delta\Delta G_\text{hyd}$ are used to compare different free-energy methods against each other and against experimental values (Figure \ref{fig: thermodynamic_cycle}). For three molecules $i$, $j$, and $k$, it holds that\cite{Jorgensen1988}
\begin{alignat}{4}
\Delta\Delta G_\text{hyd}^{ji} &= \Delta G_\text{hyd}^j - \Delta G_\text{hyd}^i &= \Delta G_\text{wat}^{ji} - \Delta G_\text{vac}^{ji}\\
\Delta\Delta G_\text{hyd}^{ki} &= \Delta G_\text{hyd}^k - \Delta G_\text{hyd}^i &= \Delta G_\text{wat}^{ki} - \Delta G_\text{vac}^{ki}\\
\Delta\Delta G_\text{hyd}^{kj} &= \Delta \Delta G_\text{hyd}^{ki}-\Delta\Delta G_\text{hyd}^{ji} \,,
\end{alignat}
where $\Delta G_\text{hyd}^i$ is the hydration free energy of molecule $i$, $\Delta G_\text{vac}^{ji}$ is the free-energy difference between molecules $i$ and $j$ in vacuum, $\Delta G_\text{wat}^{ji}$ is the free-energy difference between the two molecules in water, and $\Delta\Delta G_{hyd}^{ji}$ is the hydration free-energy difference between the two molecules (relative hydration free energy).

In classical MD simulations, hydration free energies are typically calculated with so-called alchemical free-energy methods.\cite{Shirts2007,Christ2009,Mobley2014,Hansen2014} Such methods transform a molecule (or its interaction with the environment) into another one \textit{via} nonphysical pathways.\cite{Mobley2014} The following sections give a brief overview of the two free-energy methods used in the present study.

\subsubsection{Thermodynamic Integration (TI)}
TI is a well-established method to calculate free-energy differences.\cite{Kirkwood1935} For two end-states $A$ and $B$, and using a linear coupling scheme, the potential energy of the system is defined as,
\begin{align}
    V(\mathbf{r}; \lambda) = (1-\lambda) ~ V_A(\mathbf{r}) + \lambda ~ V_B(\mathbf{r}) \, .
\end{align}
At $\lambda=0$ and $\lambda=1$, the potential energy corresponds to that of end-state $A$ and end-state $B$, respectively. This defines a $\lambda$-dependent path between the two end-states. After carrying out independent simulations at discrete $\lambda$-values between 0 and 1, the free-energy difference between states $A$ and $B$ can be estimated as\cite{Kirkwood1935}
\begin{align}
    \Delta G_{BA} = \int\limits^{1}_{0} \left< \frac{\partial V(\lambda)}{\partial \lambda} \right>_{\lambda} \,\text{d}\lambda \, .
\end{align}

\subsubsection{Replica-Exchange Enveloping Distribution Sampling (RE-EDS)}
RE-EDS is a multistate free-energy method,\cite{Sidler2016,Sidler2017,Ries2021} which combines Hamiltonian replica exchange (RE)\cite{Hansmann1997,Sugita2000} with enveloping distribution sampling (EDS).\cite{Christ2007,Christ2008} In EDS, a reference state $V_R$ is defined based on $N$ end-states as
\begin{align}
    V_R\left(\mathbf{r};s,\mathbf{E}^R\right) = -\frac{1}{\beta s}\ln\left[\sum\limits_{i=1}^N e^{-\beta s\left(V_i(\mathbf{r})-E_i^R\right)}\right] \, ,
\end{align}
where $s$ is the smoothness parameter ($s>0$), $\mathbf E^R$ is a vector of energy offsets, and $\beta=1/(k_B~T)$. 

At high $s$-values ($\sim$~1.0), due to the negative exponent, the end-state with the lowest value of $V_i(\mathbf{r}) - E_i^R$ contributes the most to the sampling of $V_R$. As $s$ decreases, it ``smoothes'' the potential-energy landscape such that all end-states contribute to the reference state, leading to an unphysical intermediate situation,\cite{Sidler2016} referred to as ``undersampling''.\cite{Riniker2011} The energy offsets control the contribution of the end-states to the reference-state potential. Optimal energy offsets ensure equal weights of all end-states in the reference state.

The free-energy difference between any pair of end-states can then be obtained from a single simulation as\cite{Christ2007,Christ2008}
\begin{align}
    \Delta G_{BA} = -\frac{1}{\beta}\ln\frac{\langle e^{-\beta(V_B-V_R)}\rangle_R}{\langle e^{-\beta(V_A-V_R}\rangle_R} \, .
\end{align}
For EDS simulations, it is essential to choose an optimal $s$-value along with optimal energy offsets to achieve adequate sampling of all end-states, which is non-trivial for more than two end-states.\cite{Riniker2011} RE-EDS enhances the sampling and simplifies the parameter optimization by simulating several replicas with different $s$-values in parallel, and performing replica exchanges between them.\cite{Sidler2016}

In recent studies, RE-EDS was successfully used to calculate relative binding and hydration free energies for molecules containing relatively large structural changes (i.e. R-group modifications and core-hopping transformations such as ring opening/closing and ring size changes).\cite{Ries2021,Ries2022}

%%%%%%%%%%%% METHODS %%%%%%%%%%%%%%
\section{Methods}

\subsection{Implementation of amber2gromos}

The \textit{amber2gromos} program is a novel C++\cite{Stroustrup1995} tool integrated into the GROMOS++ package of programs.\cite{Eichenberger2011} 
It converts AMBER topologies to a GROMOS-compatible file format. For a given AMBER topology, the program parses the input topology file, converts the force-field parameters, and outputs a GROMOS topology. The conversion steps of the program are outlined in the Supporting Information. An example of an AMBER topology for aniline, and the resulting GROMOS topology translated with \textit{amber2gromos} can be found in the Supporting Information in Listings 1 and 2, respectively.%%SUPPL lst: ambtop , lst:grotop

%% changes to gromos source code
The conversion process described in the Supporting Information generates a valid GROMOS topology from a given AMBER topology. However, there is still a difference between AMBER and GROMOS in the handling of the 1,4-electrostatic interactions (i.e., scaling by a factor $1/1.2$\cite{Wang2004,AMBER_manual_2016} in AMBER). As 1,4-electrostatic interactions are not scaled in the GROMOS force fields, the scaling option is not supported within a GROMOS topology. Therefore, some changes had to be made to the GROMOS\cite{Schmid2012} source code. A new block type ``AMBER'' was added to the GROMOS input file, which contains a switch (0 = off,  1 = on) for the scaling of the electrostatic third-neighbor interactions together with the scaling parameter (e.g., $1.2$, for scaling by $1/1.2$). When the switch is off, the scaling parameter is set to 1.0 so that no scaling is applied.
When a reaction field (RF) correction\cite{Tironi1995} is used in GROMOS to account for long-range electrostatic interactions, these interactions are calculated as,\cite{Riniker2018,GROMOS_manual_v2}
\begin{align}
    V^\text{ele} &= \sum\limits_i\sum\limits_{j>i}\frac{q_i q_j}{4\pi\varepsilon_0}\left[\frac{1}{r_{ij}}-\frac{C_{RF}r_{ij}^2}{2R_{RF}^3}-\frac{1-0.5C_{RF}}{R_{RF}}\right] \, ,
\end{align}
where $q_i$ and $q_j$ are the charges of atoms $i$ and $j$, $\varepsilon_0$ is the permittivity of vacuum, and $R_{RF}$ is the cutoff distance.\cite{Tironi1995,Riniker2018,GROMOS_manual_v2,GROMOS_manual_v6} Here, $C_{RF}$ is a constant characterizing the effect of the RF continuum, given by\cite{Tironi1995,GROMOS_manual_v2}
\begin{align}
C_{RF} = \frac{(2\varepsilon_{cs}-2\varepsilon_{RF})(1+\kappa_{RF}R_{RF})-\varepsilon_{RF}(\kappa_{RF}~R_{RF})^2}{(\varepsilon_{cs}+2\varepsilon_{RF})(1+\kappa_{RF}R_{RF})+\varepsilon_{RF}(\kappa_{RF}~R_{RF})^2}\,,
\end{align}
where $\varepsilon_{cs}$ is the relative permittivity of the medium in which the simulation is performed, $\varepsilon_{RF}$ is the RF permittivity, and $\kappa_{RF}$ is the inverse Debye screening length.\cite{GROMOS_manual_v2} The electrostatic interactions in the GROMOS\cite{Christen2005,Schmid2012} source code were changed to
\begin{align}
    V^\text{ele} &= \sum\limits_i\sum\limits_{j>i}\frac{q_i q_j}{4\pi\epsilon_0}\left[\frac{s_{ij}}{r_{ij}}-\frac{C_{RF}r_{ij}^2}{2R_{RF}^3}-\frac{1-0.5C_{RF}}{R_{RF}}\right] \, ,
\end{align}
where $s_{ij}$ corresponds to the AMBER scaling factor when $i$ and $j$ are third neighbors, and 1.0 otherwise. Note that in this expression, only the direct Coulombic interactions are scaled.\cite{Bergdorf2003, Christen2005}

%% charge groups
As the AMBER force fields do not use charge groups, by default, all the atoms in a molecule/residue are assigned to the same charge group for the simulations with RF electrostatics\cite{Tironi1995} in GROMOS (which is only appropriate for small molecules). In addition, there is also the option to consider each atom as defining its own charge group (resulting in an atom-based cutoff). The user can also prepare a so-called charge-group file, defining which (consecutive) atoms should be assigned to the same charge group. There are currently ongoing efforts to combine RE-EDS with the shifting-function based scheme developed by Kubinc\'ova \textit{et al.}\cite{Kubincova2020} such that an atom-based cutoff can be employed with RF electrostatics without cutoff artefacts. This will become the default choice when using the AMBER (and OpenFF) force fields in GROMOS in the near future, mitigating the requirement to define charge groups altogether.

\subsection{RE-EDS Pipeline}
Recently, Ries \textit{et al.}\cite{Ries2021} presented an improved pipeline to perform relative free-energy calculations with RE-EDS. The RE-EDS pipeline can be divided into three main phases: parameter exploration, parameter optimization, and production. The parameter exploration phase is used to generate relevant configurations for all end-states, to determine a lower bound for the $s$-values that ensures undersampling, and to obtain initial estimates for the energy offsets. During the parameter optimization, the distribution of $s$-values and the energy offsets $E_i^R$ are refined such that all end-states are sampled nearly equally at $s=1$. Finally, a production run is performed to calculate the free-energy differences between all pairs of end-states simultaneously. The entire workflow can be executed using the Python3\cite{VanRossum2009} \textit{reeds} module.\cite{Ries2021reeds} In this work, the RE-EDS pipeline was applied to the calculation of relative hydration free energies of benzene derivatives.

\subsection{Datasets}
The main goals of this work were to validate the topology conversion with \textit{amber2gromos} and the accompanying changes in the GROMOS MD engine, as well as to investigate the performance of RE-EDS for systems with many end-states. To this end, two sets of benzene derivatives with experimental hydration free energies in the FreeSolv\cite{Mobley2014,Duarte2017} database  were selected. This test system was chosen due to the common benzene core, the relatively small size (which limits deviations due to sampling issues), and the availability of calculated and experimental reference data. The potential of RE-EDS to handle larger perturbations has already been demonstrated in previous publications.\cite{Ries2021,Ries2022}.

\subsubsection{Set A: Six Benzene Derivatives}
\begin{figure}[h]
    \centering
    \includegraphics[width=0.6\textwidth]{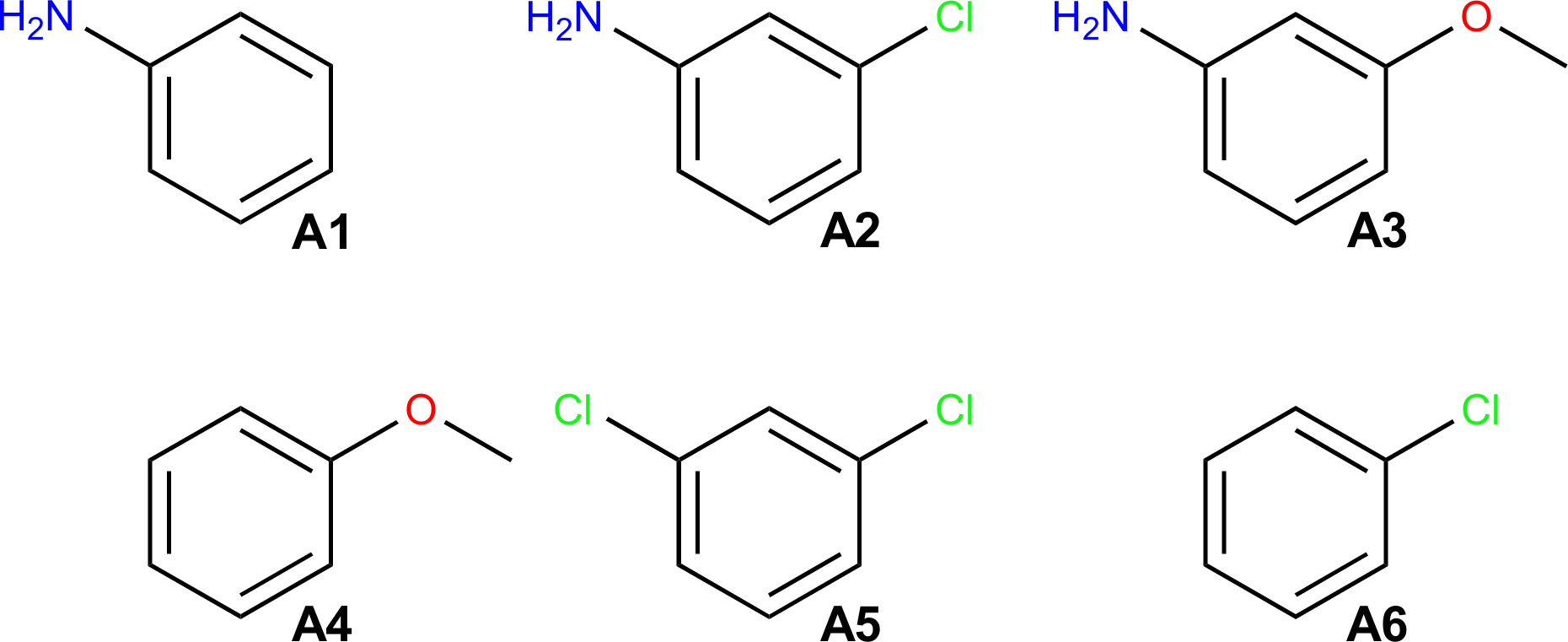}
    \caption{Set A consists of six benzene derivatives, selected from the FreeSolv\cite{Mobley2014,Duarte2017} database. A table with the molecule indices, the FreeSolv identifiers, the SMILES strings, and the names of the molecules can be found in Table S1 in the Supporting Information.} %%SUPPL tab: list_setA
    \label{fig: setA}
\end{figure}

As a first test set, six benzene derivatives were selected (Figure \ref{fig: setA}). The number of end-states was deliberately kept small to efficiently test the implementation. The mol2 and frcmod files provided by FreeSolv\cite{Mobley2014,Duarte2017} were used to generate AMBER topologies using \textit{tleap} (AmberTools16).\cite{AMBER_manual_2016} The topologies were then converted to GROMOS format using \textit{amber2gromos}, and to GROMACS\cite{Berendsen1995,Abraham2015} format using ParmEd.\cite{Swails2010} The force-field parameters of the original AMBER topologies and of the generated GROMOS/GROMACS topologies were compared manually. In addition, a single energy evaluation was performed for the individual molecules in vacuum/water using both GROMOS and GROMACS. It was verified that the covalent and non-bonded energy terms calculated by the two different MD engines were nearly identical. Longer MD simulations of 5~ns were then carried out for each molecule in vacuum to compare properties such as the system temperature and the different energy terms.

Next, the free-energy differences in vacuum/water were calculated for all 15 molecule pairs from RE-EDS simulations containing six end-states. Complementary single-topology pairwise TI calculations were performed using GROMACS. The relative hydration free energies $\Delta\Delta G_\text{hyd}^{ji} = \Delta G_\text{wat}^{ji} - \Delta G_\text{vac}^{ji}$ were calculated from the RE-EDS and TI calculations. They were compared to the relative hydration free energies $\Delta\Delta G_\text{hyd}^{ji} = \Delta G_\text{hyd}^{j} - \Delta G_\text{hyd}^{i}$ obtained from the experimental and calculated hydration free energies reported in the FreeSolv\cite{Mobley2014,Duarte2017} database.

\subsubsection{Set B: 28 Benzene Derivatives}
\begin{figure}[h]
    \centering
    \includegraphics[width=0.9\textwidth]{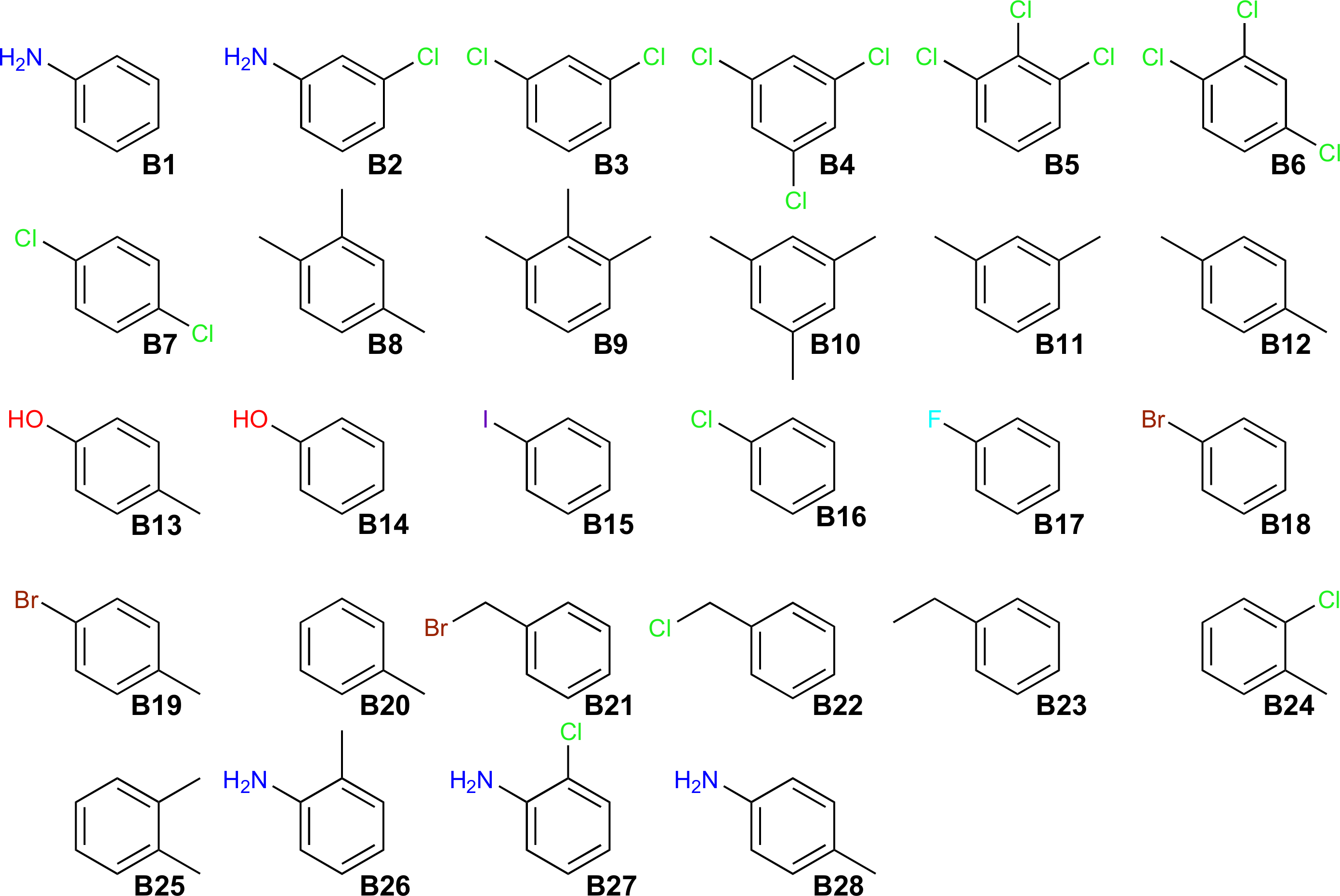}
    \caption{Set B consists of 28 benzene derivatives, selected from the FreeSolv\cite{Mobley2014,Duarte2017} database. Set B was further subdivided into subset Ba (B1 - B14) and subset Bb (B1 along with B15 - B28). A table with the molecule indices, the FreeSolv identifiers, the SMILES strings, and the name of the molecules can be found in Table S2 in the Supporting Information.} %%SUPPL tab: list_setB
    \label{fig: setB}
\end{figure}

To investigate the performance of RE-EDS for a larger number of end-states, the set of benzene derivatives was increased to 28 (Figure \ref{fig: setB}). This extended test set serves as a proof of principle that RE-EDS can be used to calculate free-energy differences for systems with larger numbers of end-states. It is currently the largest set of end-states considered in a RE-EDS simulation. Previous studies involved systems with five\cite{Ries2021}, six\cite{Ries2022}, nine\cite{Sidler2017}, and ten\cite{Sidler2016,Ries2022} end-states.
The free-energy differences were calculated for all 378 molecule pairs in vacuum/water from RE-EDS simulations containing 28 end-states. Analogously to set A, the relative hydration free energies from the RE-EDS calculation were compared to the hydration free energies reported in the FreeSolv\cite{Mobley2014,Duarte2017} database.

To further investigate the performance of RE-EDS for such a large set of end-states, set B was subdivided into two smaller subsets, Ba and Bb. Subset Ba consisted of molecules B1 - B14, and subset Bb of molecule B1 together with molecules B15 - B28. For the two subsets, RE-EDS simulations were carried out in vacuum/water to calculate the relative hydration free energies for all end-state pairs in both sets. The relative hydration free energies between the molecule pairs $j$-$k$ that were not in the same subset were calculated \textit{via} molecule B1, which was present in both subsets, as
\begin{align}
    \Delta\Delta G_\text{hyd}^{kj} &= \Delta\Delta G_\text{hyd}^{k,B1}-\Delta\Delta G_\text{hyd}^{j,B1}
\end{align}

\subsection{Simulation Details}

\begin{figure}[h]
    \centering
    \includegraphics[width=0.6\textwidth]{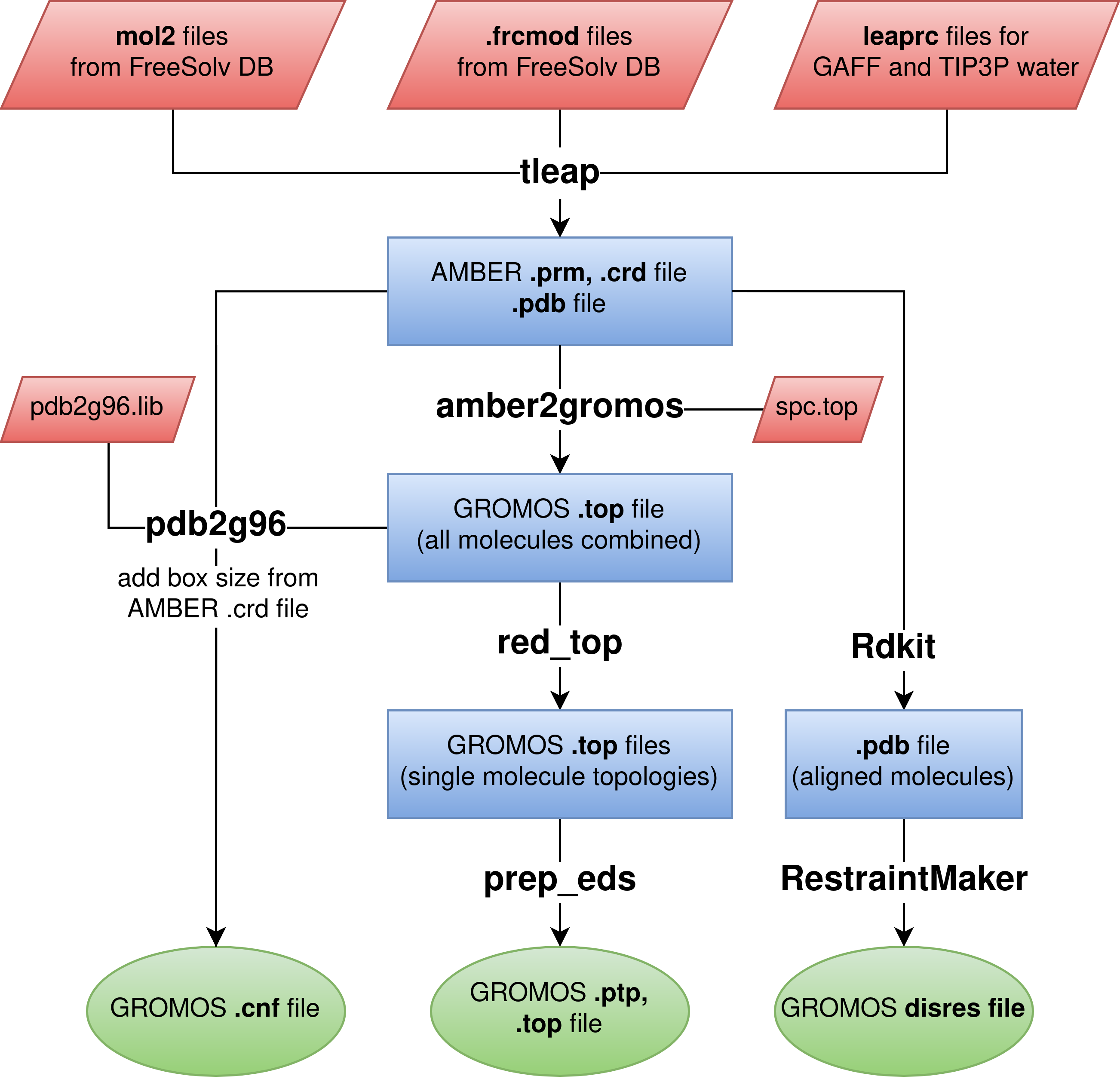}
    \caption{Schematic illustration of the RE-EDS input file preparation. The input files (topology, perturbed topology, coordinates and distance restraints) for the (RE-)EDS simulations in GROMOS were created from the frcmod and mol2 files of the FreeSolv\cite{Mobley2014,Duarte2017} database. This workflow can easily be extended to also perform the molecule parameterization (i.e., to generate mol2 and frcmod files) using \textit{antechamber} and \textit{parmchk}.\cite{Wang2001,Wang2004,Wang2006}}
    \label{fig: gaff_workflow}
\end{figure}

%% preparation of input files
The AMBER topologies were generated using AmberTools16\cite{AMBER_manual_2016} and the GAFF 1.7\cite{Wang2004} force field with the mol2 and frcmod files provided in the FreeSolv\cite{Mobley2014,Duarte2017} database as a starting point. The atomic charges were generated using the AM1-BCC\cite{Jakalian2000,Jakalian2002} approach. The input files for the single-topology TI calculations in GROMACS were prepared with FESetup.\cite{Loeffler2015} The input files for the GROMOS RE-EDS simulations were prepared using \textit{amber2gromos} as well as the GROMOS++\cite{Eichenberger2011} programs \textit{pdb2g96}, \textit{red\_top}, and \textit{prep\_eds}. The molecules were aligned for the RE-EDS simulations using the RDKit\cite{Rdkit} (details in Figures S2 and S8 in the Supporting Information). %%SUPPL fig: alignment_setB, fig: alignment_setA
Since the coordinates of all molecules are present separately in the system, the molecules are in principle able to drift away from each other during a simulation. To ensure that the molecules remain well-aligned during the whole simulation, atomic distance restraints were applied. The pairwise distance restraints were generated with \textit{RestraintMaker}.\cite{Ries2022} \textit{RestraintMaker} chooses reference distances $r_0$ between restrained atoms according to the input alignment. For some molecule pairs, the ring atoms did not perfectly overlap in the initial alignment. The reference distances assigned by \textit{RestraintMaker} were manually set to 0 for those pairs. Four atoms of each molecule were restrained to four atoms of two other molecules, forming a chain of pairwise distance restraints. For set B with 28 molecules, the chain arrangement allowed for relatively large deviations between the molecules furthest apart in the chain. Therefore, additional distance restraints were manually added for four molecule pairs on opposite sides of the chain. The workflow to generate the input files for (RE-)EDS simulations with GAFF\cite{Wang2004} parameters is illustrated in Figure \ref{fig: gaff_workflow}.

%%simulation parameters common to all simulations
The RE-EDS simulations were performed with a modified version of GROMOS\cite{Schmid2012,gromos_net} 1.5.0 and the open-source Python3\cite{VanRossum2009} \textit{reeds} module.\cite{Ries2021reeds} The TI simulations were performed in GROMACS\cite{Berendsen1995,Abraham2015} version 2016.6. For the simulations in water, the TIP3P water model\cite{Jorgensen1983} was used. A single cutoff radius of 1.2~nm was used for the calculation of the non-bonded interactions. The integration timestep was set to 2~fs and the pairlist was updated every five steps. Long-range non-bonded interactions were calculated in GROMOS using a reaction-field correction\cite{Tironi1995} with $\varepsilon_{\text{RF}}=1$ for the simulations in vacuum and $\varepsilon_{\text{RF}}=78.5$ for the simulations in water.\cite{Glattli2002,Riniker2011b} In GROMACS, the long-range non-bonded interactions were calculated using a plain cut-off in vacuum, and  smooth particle mesh Ewald (SPME)\cite{Ewald1921,Darden1993,Essmann1995} in water with a grid spacing of 0.1~nm and an interpolation of order 6%, analogous to the FreeSolv setup
. To maintain the temperature at 298.15~K and the pressure at 0.06102~kJ~mol$^{-1}$~nm$^{-3}$~($\approx~$1~atm), Berendsen thermostat and barostat\cite{Berendsen1984} were used in GROMOS for the simulations in water. For the GROMACS TI calculations and the RE-EDS calculations in vacuum, the leap-frog stochastic dynamics integrator was used, so that no temperature scaling was necessary. In the TI calculations, the pressure in water was kept constant at 1.01325~bar~($\approx$~1~atm) using a Parrinello-Rahman barostat.\cite{Parrinello1980} All bonds were constrained with the LINCS algorithm \cite{Hess1997} (for the TI calculations, 12$^{th}$ order) or the SHAKE algorithm\cite{Ryckaert1977} (for the RE-EDS calculations, relative tolerance 10$^{-4}$), respectively, and harmonic bond-angle bending was employed.
In the RE-EDS simulations in GROMOS, the force constant for the distance restraints was set to $5000$ kJ$/($mol$\cdot$nm$^2)$.\cite{Ries2022}

%%FreeSolv simulation parameters and different pathways of methods
The calculated hydration free energies reported in the FreeSolv\cite{Mobley2014,Duarte2017} database were obtained from alchemical MBAR\cite{Shirts2008} simulations in GROMACS with 20 $\lambda$-values, each 5~ns long. In the first five intermediate states, the electrostatic interactions were modified, and in the last 15 states, the Lennard-Jones terms were changed.\cite{Duarte2017} In this work, the relative hydration free energies for set A were obtained from pairwise TI simulations and from RE-EDS simulations in vacuum/water. For set B, they were determined from RE-EDS simulations in vacuum/water. Here, it is important to note that the three methods used different pathways of the thermodynamic cycle to calculate the relative hydration free energies. Considering Figure \ref{fig: thermodynamic_cycle}, the MBAR calculations\cite{Mobley2014,Duarte2017} correspond to the yellow pathways, the TI calculations to the pink pathways, and the multistate RE-EDS calculations to the blue pathways.

%%data availability
The input files for the RE-EDS simulations can be found at \url{https://github.com/rinikerlab/reeds/tree/main/examples/systems/benzenes_amber2gromos}.

\subsubsection{Set A}
To obtain the pairwise free-energy differences in vacuum/water with TI calculations in GROMACS, 21 $\lambda$-values were used in vacuum and 27 $\lambda$-values in water. In vacuum they were spread in steps of 0.05 between 0 and 1. In water, they were spread in steps of 0.05 between 0.1 and 0.9, and more densely in steps of 0.02 around the extreme values (i.e., between 0 and 0.1, and between 0.9 and 1). The values were spread more densely around the extreme values for the simulations in water to smooth discontinuities in the $\langle\partial V/\partial\lambda\rangle$-curves. Such discontinuities were mainly observed between 0.0 and 0.1, but also occurred between 0.9 and 1.0 for molecule pairs A1 - A2, A1 - A3, and A2 - A3 (Figure S3 in the Supporting Information). %%SUPPL
The masses of the end-state atoms were not perturbed, but kept to the masses of the first end-state. After a short steepest-descent energy minimization with maximally 5000 steps, the systems were equilibrated for 0.5~ns.  The free-energy differences were calculated from five independent production runs in vacuum/water. The production runs were 5~ns long at each $\lambda$-point.

To generate relevant configurations of all end-states for the RE-EDS calculations, six EDS simulations in vacuum/water of 2~ns length were carried out at $s=1$. The energy offsets were biased towards one of the end-states (molecules) in each of the simulations by setting one energy offset to 500~kJ~mol$^{-1}$ and the others to -500~kJ~mol$^{-1}$. The optimized configurations were used for the starting-state mixing (SSM) approach.\cite{Ries2021} Simultaneously, 21 EDS simulations of 0.2~ns length with $s$-values logarithmically distributed between 1 and $10^{-5}$ were used to determine the lower bound of $s$ for the RE-EDS simulations, which were set to 0.0178 in vacuum and 0.01 in water. RE-EDS simulations of 0.8~ns length were carried out to estimate the energy offsets with 11 replicas in vacuum and 12 replicas in water. Next, the $s$-distribution was optimized to achieve frequent round trips. Following the SSM approach,\cite{Ries2021} the $s$-values were re-distributed to include a replica with $s=1$ and optimized initial coordinates for each end-state (in total 12 replicas in vacuum, 13 in water). In vacuum, four replicas were added after one $s$-optimization step of 0.5~ns length. In water, two $s$-optimization steps of 0.5~ns and 1.0~ns lengths, respectively, were used, adding in total eight replicas. To achieve good sampling of all end-states, the initial energy offsets in water were rebalanced over two 0.5~ns RE-EDS simulations. 
%In vacuum, no rebalancing was required.
Finally, the free-energy differences were calculated from five independent production runs of 0.5~ns in vacuum (11 replicas) and water (16 replicas). For the production runs, the additional replicas with $s=1$, which were added during the optimization phase, were removed again.

\subsubsection{Set B}
For set B, 28 EDS simulations of 2~ns length were carried out to generate optimized configurations. The setup for the lower bound search was analogous to set A, and the values were set to 0.01 in vacuum and 0.0056 in water. For the energy offset estimation, 34 replicas were used in vacuum and 35 in water with 0.8~ns length. The $s$-optimization steps were analogous to set A, adding four replicas in vacuum and eight replicas in water. For set B, rebalancing was also required in vacuum. Both in vacuum and in water, four rebalancing steps were used, of 0.5~ns length each. The production run was 1~ns long in vacuum (34 replicas) and 2~ns long in water (39 replicas). As for set A, the obtained free-energy differences were averaged over five independent production runs.

To generate optimized coordinates, 14 EDS simulations were performed for subset Ba and 15 EDS simulations for subset Bb. The lower bounds for $s$ were 0.01 for subset Ba in vacuum, 0.032 for subset Bb in vacuum, and 0.01 for both subsets in water. For the energy offset estimation, 20 (Ba, vacuum), 19 (Bb, vacuum), 20 (Ba, water), and 21 (Bb, water) replicas were used. For both subsets in vacuum, one $s$-optimization step was sufficient. For subset Ba in water, two $s$-optimization steps of 0.5~ns and 1.0~ns, respectively, were necessary, while for subset Bb in water, only one $s$-optimization step was needed. In vacuum, three energy offset rebalancing steps were carried out for both subsets. In water, four rebalancing steps were used for subset Ba, and three for subset Bb. Finally, the production runs were 1~ns long in vacuum (20 replicas for Ba, 19 replicas for Bb) and 2~ns long in water (24 replicas for Ba, 21 replicas for Bb). Again, the obtained free-energy differences were averaged over five independent production runs.

\subsection{Analysis}
The analysis of the simulations was carried out using GROMOS++ \cite{Eichenberger2011} and PyGromosTools.\cite{Lehner2021} The following Python packages were used for visualization and analysis:
Matplotlib,\cite{Hunter2007}
mpmath,\cite{Johansson2013}
NumPy,\cite{VanDerWalt2011}
Pandas,\cite{Mckinney2010}
SciPy,\cite{Virtanen2020}
and Seaborn.\cite{Waskom2021}
For all sets/subsets, the root-mean-square error (RMSE), the mean absolute error (MAE) and the Spearman\cite{Spearman1904} correlation coefficient were calculated between the different simulation methods and the experimental values.

%%%%%%%%%%%% RESULTS %%%%%%%%%%%%%%
\section{Results}

\subsection{Validation of amber2gromos}
For the molecules of set A, the manual topology comparison showed that the topologies generated by \textit{amber2gromos} were almost identical to the GROMACS topologies generated by ParmEd\cite{Swails2010} (apart from differences in units, functional forms, and slight numerical differences). Apart from negligible numerical differences, the potential-energy terms of the 0$^{th}$ integration step in vacuum and water were identical for the simulations in GROMOS and in GROMACS.

\begin{figure}[h]
    \centering
    \includegraphics[width=0.95\textwidth]{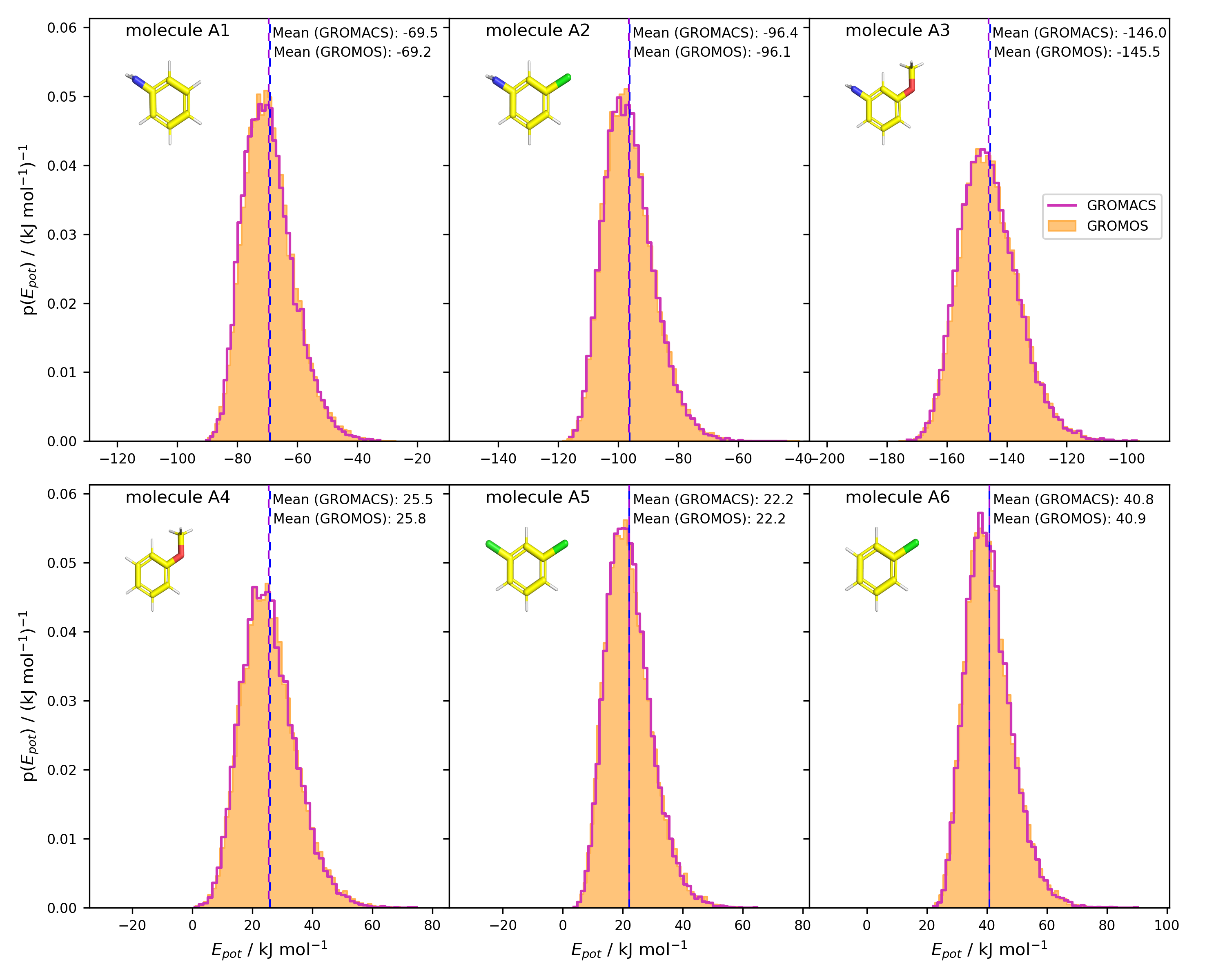}
    \caption{Potential-energy distributions of single-molecule simulations of set A based on 5~ns vacuum simulations in GROMOS\cite{Schmid2012} (orange bars) and GROMACS\cite{Abraham2015} (pink lines). The topologies for the simulations are based on the AMBER topologies taken from the FreeSolv\cite{Mobley2014,Duarte2017} database. The GROMACS topologies were translated with ParmEd\cite{Swails2010} and the GROMOS topologies were converted with \textit{amber2gromos}. Energies were written every 100 timesteps (i.e., every 200~fs), and the first 1.25~ns of the simulations were discarded as equilibration.}
    \label{fig: single_vacuum_e_pot}
\end{figure}

After this initial validation, simulations of 5~ns length were performed in vacuum. The distributions of different energy terms as well as the system temperature were compared. They were all qualitatively similar, with almost identical mean values (Figure \ref{fig: single_vacuum_e_pot} and Figure S1 in the Supporting Information). 

\subsection{Calculation of Relative Hydration Free Energies for Set A}
For set A, the 15 pairwise free-energy differences in vacuum/water obtained from the RE-EDS calculations were first compared to the ones from the TI calculations in GROMACS. With an RMSE of 1.9~kJ~mol$^{-1}$ (vacuum) and 2.0~kJ~mol$^{-1}$ (water) and a MAE of 1.6~kJ~mol$^{-1}$ (vacuum) and 1.7~kJ~mol$^{-1}$ (water), the results agreed well. The Spearman correlation coefficent was 1.00 for the vacuum and the water simulations (Figure S4 in Supporting Information).

\begin{figure}[h]
    \centering
    \includegraphics[width=0.45\textwidth]{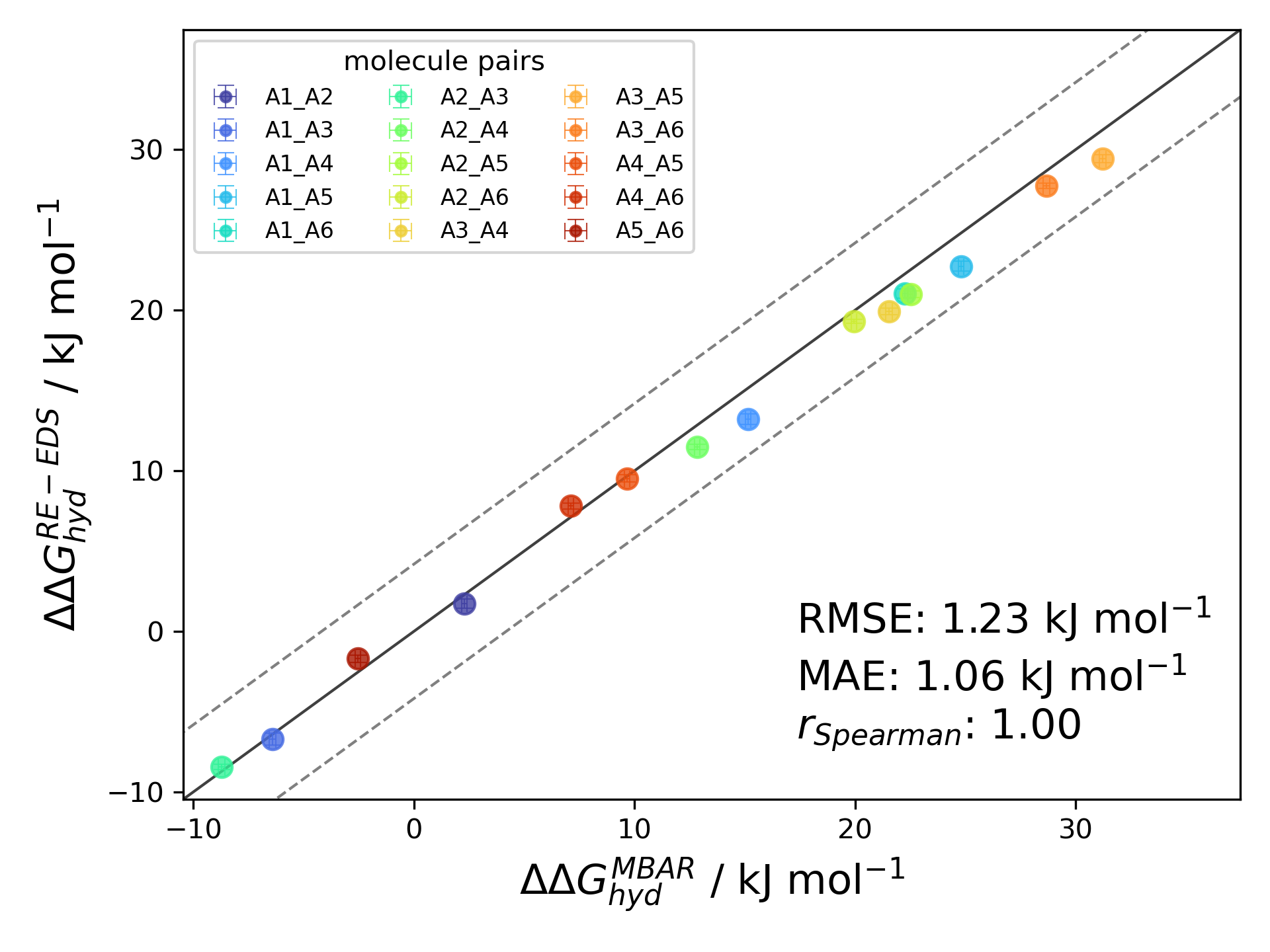}
    \includegraphics[width=0.45\textwidth]{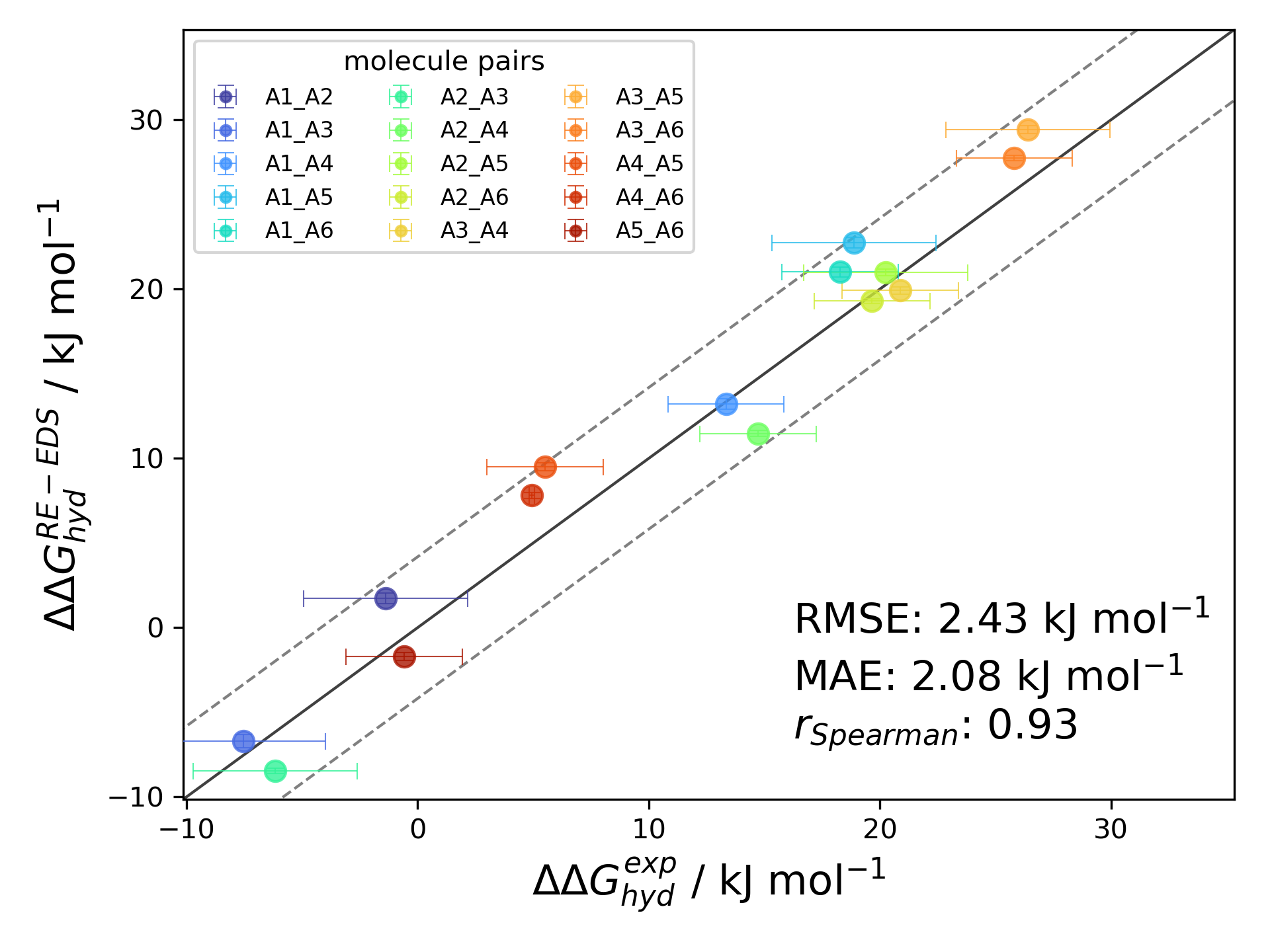}
    \caption{Comparison of the relative hydration free energies of set A: RE-EDS ($\Delta\Delta G_\text{hyd}^\text{RE-EDS}$) versus MBAR ($\Delta\Delta G_\text{hyd}^\text{MBAR}$) (left), and RE-EDS versus experiment ($\Delta\Delta G_\text{hyd}^\text{exp}$) (right) as reported by the FreeSolv\cite{Mobley2014,Duarte2017} database. The gray diagonal lines correspond to perfect alignment within $\pm$~4.184~kJ~mol$^{-1}$ ($\pm$~1~kcal~mol$^{-1}$). The RE-EDS results were averaged over five independent production runs in vacuum/water and the errors of the $\Delta G$ values correspond to the standard deviation over the five repeats. The error estimate of the $\Delta \Delta G$ values was calculated \textit{via} Gaussian error propagation. The numerical values are provided in Table S3 in the Supporting Information. A plot of the deviations from experiment for the different methods is shown in Figure S7 in the Supporting Information. All pairwise comparisons between the different simulation methods and the experimental results are provided in Figure S5.} 
    \label{fig: ddg_setA}
\end{figure}

Next, the relative hydration free energies from the different sources (i.e., TI in GROMACS, RE-EDS in GROMOS, MBAR in GROMACS\cite{Duarte2017}, and experimental\cite{Mobley2014,Duarte2017}) were compared. The results from all three simulation methods agreed well with the other calculated results, as well as with the experimental values (Figure \ref{fig: ddg_setA} and Table \ref{tab: summary_setA}). The RMSEs against the experimental results were 2.4~kJ~mol$^{-1}$ (TI), 2.4~kJ~mol$^{-1}$ (RE-EDS), and 3.1~kJ~mol$^{-1}$ (MBAR). The corresponding MAEs were 2.0~kJ~mol$^{-1}$ (TI), 2.1~kJ~mol$^{-1}$ (RE-EDS), and 2.7~kJ~mol$^{-1}$ (MBAR), respectively. The calculated relative hydration free energies also had a high correlation with the experimental results, with Spearman correlation coefficients ranging between 0.93 and 0.94. The full details are provided in  Table S3 in the Supporting Information. 

While all three free-energy methods achieve comparable and accurate results, the RE-EDS calculations require by far the lowest accumulated simulation time. The total simulation time (pre-processing and production) for the RE-EDS calculations was about 115~ns. For the TI calculations, the total simulation time (equilibration and production) was 3960~ns. This could of course be reduced by calculating only the minimal number of required pairwise free-energy differences (i.e., $N$-1, which is five for set A). The production time could likely also be reduced to a total of 2 - 3~ns without affecting the convergence significantly. Both measures would reduce the total simulation time required for the TI calculations to about 600 - 840~ns. However, even then, the total simulation time would still be more than five times longer than for the RE-EDS calculations. The calculated values reported in FreeSolv\cite{Mobley2014,Duarte2017} required 618~ns total simulation time (equilibration and production). Also here, the simulation time was chosen with convergence in mind, and could probably be reduced. However, even with 2 - 3~ns production runs, the total simulation time would still be about 2 - 3 times the simulation time required by RE-EDS. Plots of the convergence of the free-energy calculations with RE-EDS as well as the $\lambda$-curves for the TI simulations can be found in Figures S6 and S3 in the Supporting Information. 

\begin{table}[h]
\caption{Set A: Overview of statistical metrics (RMSE, MAE, and Spearman correlation coefficients) with respect to the experimental results, and total simulation time for the different free-energy methods. The RE-EDS and TI results were averaged over five independent production runs in vacuum/water and the errors of the $\Delta G$ values correspond to the standard deviation over the five repeats. The error estimate of the $\Delta \Delta G$ values was calculated \textit{via} Gaussian error propagation. 
The uncertainties of the RMSE and MAE values were estimated from the distribution of RMSE and MAE when a random selection of up to four molecules was removed from the calculations (5000 repetitions).
The accumulated simulation time is split into preparation (pre-processing, equilibration) and production time. The full table can be found in Table S3 in the Supporting Information.}
\label{tab: summary_setA}
\centering
\begin{tabular}{ | l | c | c | c | }
    \hline
    \multicolumn{1}{|c|}{} \rule{0pt}{12pt}& $\Delta\Delta G_\text{hyd}^\text{MBAR}$\cite{Mobley2014,Duarte2017} & $\Delta\Delta G_\text{hyd}^\text{TI}$ & $\Delta\Delta G_\text{hyd}^\text{RE-EDS}$\\ 
 \hline \hline
        {RMSE [kJ~mol$^{-1}$]} & $3.1~\pm~0.4$ & $2.4~\pm~0.4$ & $2.4~\pm~0.3$ \\
        {MAE [kJ~mol$^{-1}$]} & $2.7~\pm~0.3$ & $2.0~\pm~0.4$ & $2.1~\pm~0.3$ \\
        \hline
        {$r_{\text{Spearman}}$} & 0.94 & 0.93 & 0.94 \\
        \hline
        {$t_\text{preparation}$ [ns]} & 18 &  360 & 101.3 \\
        {$t_\text{production}$ [ns]} & 600 & 3600 & 13.5  \\
    \hline
\end{tabular}
\end{table}

\subsection{Calculation of Relative Hydration Free Energies for Set B}
\begin{figure}[h]
    \centering
        \includegraphics[width=0.45\textwidth]{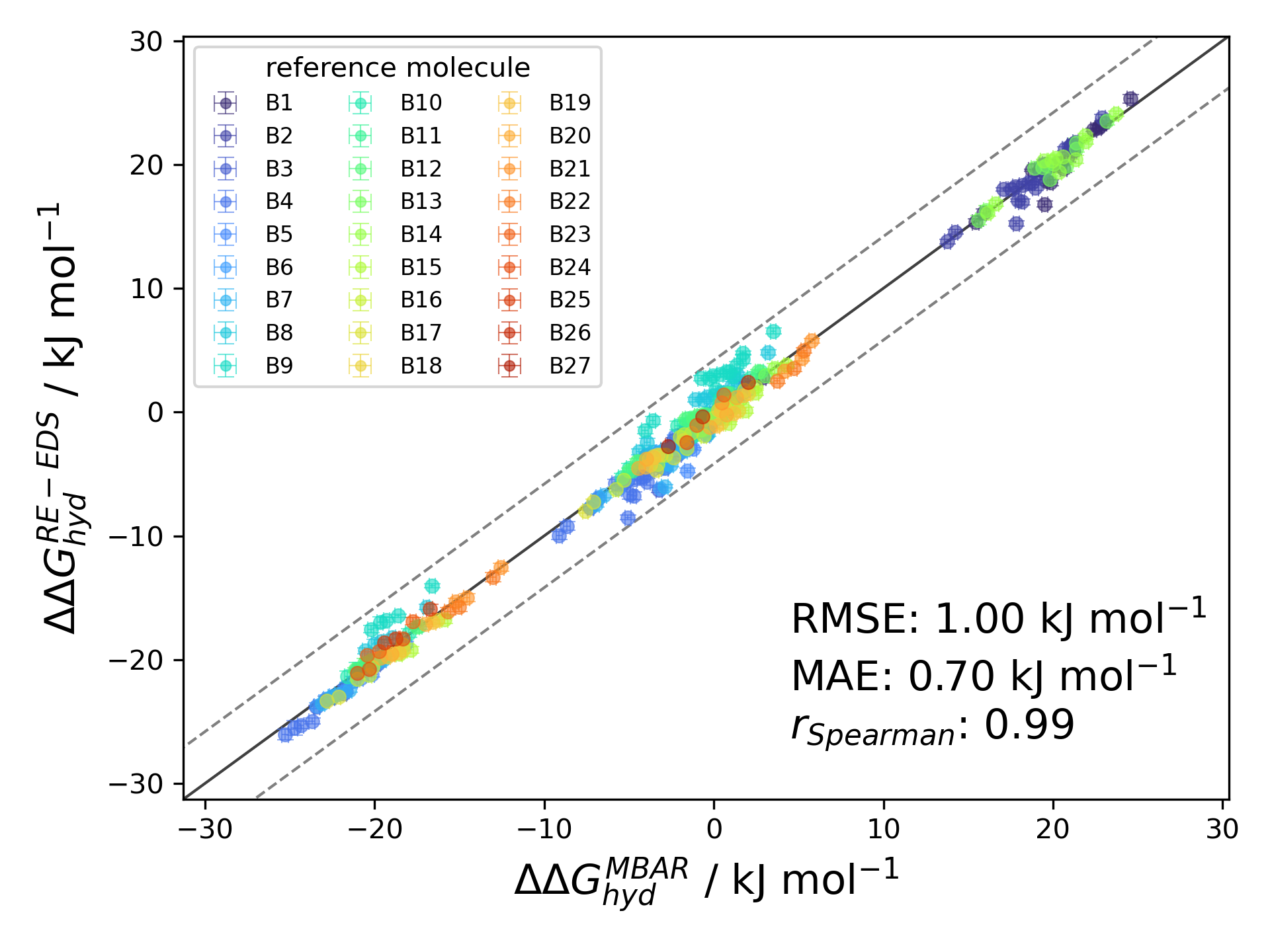}
        \includegraphics[width=0.45\textwidth]{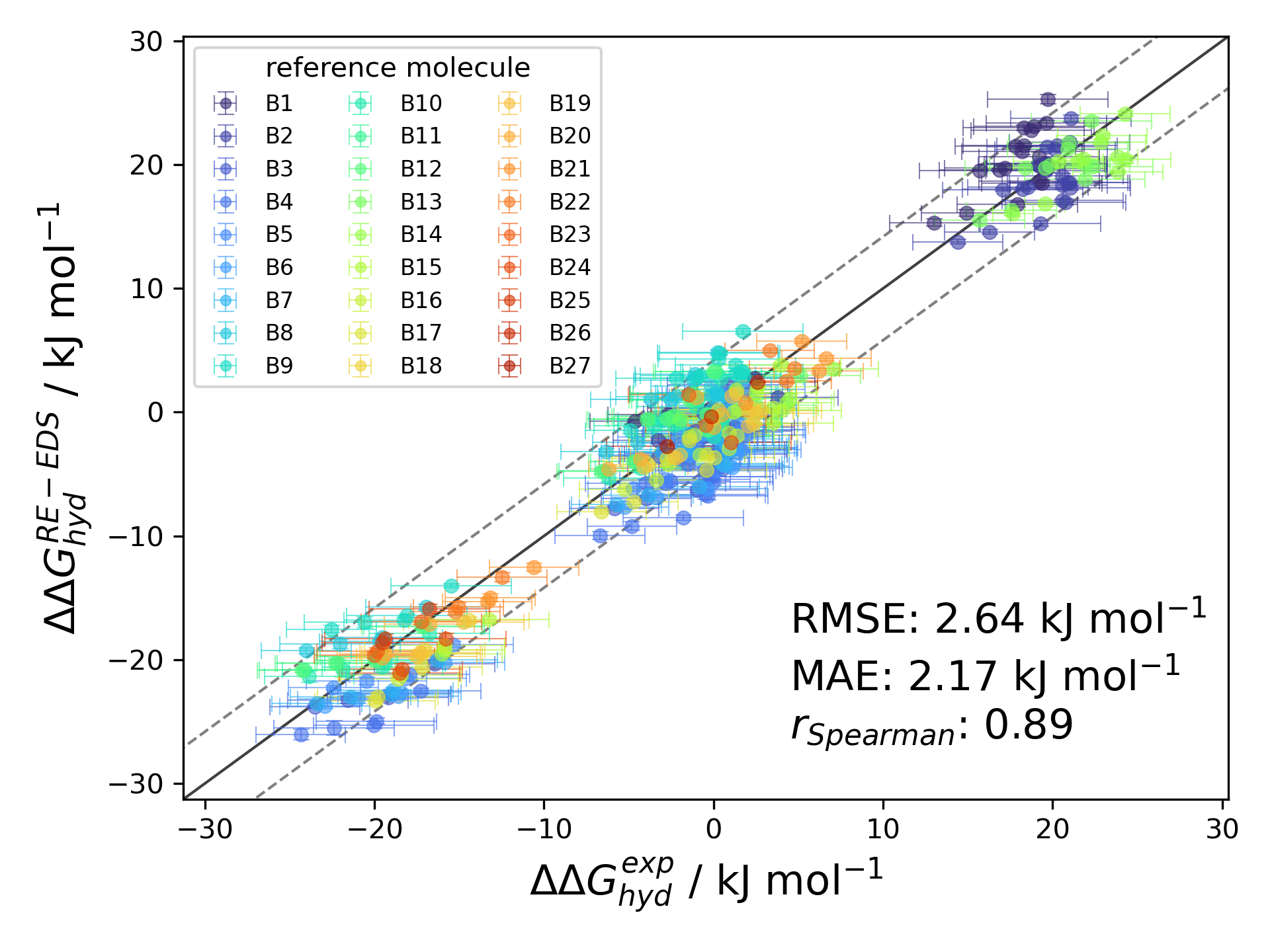}
    \caption{Comparison of the relative hydration free energies of set B: RE-EDS ($\Delta\Delta G_\text{hyd}^\text{RE-EDS}$) versus MBAR ($\Delta\Delta G_\text{hyd}^\text{MBAR}$) (left), and RE-EDS versus experiment ($\Delta\Delta G_\text{hyd}^\text{exp}$) (right) as reported by the FreeSolv\cite{Mobley2014,Duarte2017} database. The gray diagonal lines correspond to perfect alignment within $\pm$~4.184~kJ~mol$^{-1}$ ($\pm$~1~kcal~mol$^{-1}$). The $\Delta\Delta G_\text{hyd}^{ji}$ values are colored according to end-state $i$ (i.e., the ``reference molecule'' for the calculation). The RE-EDS results were averaged over five independent production runs in vacuum/water and the errors of the $\Delta G$ values correspond to the standard deviation over the five repeats. The error estimate of the $\Delta \Delta G$ values was calculated \textit{via} Gaussian error propagation. The numerical values are provided in Table S4 in the Supporting Information. A plot of the deviations from experiment for the different methods is shown in Figure S12 in the Supporting Information. All pairwise comparisons between the different simulation methods and the experimental results are provided in Figure S9.} 
    \label{fig: ddg_setB}
\end{figure}

For set B, we found again an excellent agreement between the results obtained from the RE-EDS simulations and the calculated and experimental values reported in the FreeSolv\cite{Mobley2014, Duarte2017} database (Figure \ref{fig: ddg_setB} and Table \ref{tab: summary_setB}). The RMSE against the experimental results was 2.6~kJ~mol$^{-1}$ for RE-EDS and 2.0~kJ~mol$^{-1}$ for MBAR.\cite{Mobley2014,Duarte2017} The corresponding MAEs were 2.2~kJ~mol$^{-1}$ and 1.6~kJ~mol$^{-1}$, respectively. The correlation with experiment was high for both methods, with $r_\text{Spearman}^\text{RE-EDS}=0.89$, and $r_\text{Spearman}^\text{MBAR}=0.92$. The agreement between the two simulation methods was also good, with an RMSE of 1.0~kJ~mol$^{-1}$, a MAE of 0.7~kJ~mol$^{-1}$, and a Spearman correlation coefficient of 0.99. For RE-EDS, molecule pairs B1 - B4, B4 - B8, B4 - B9, B4 - B23, B4 - B25, B6 - B8, and B6 - B9 showed deviations above 5.5~kJ~mol$^{-1}$, the largest for molecule pair B4 - B9 with 6.7~kJ~mol$^{-1}$ (Figure S12 in the Supporting Information). 
For MBAR, molecule pairs B1 - B4, B1 - B6, B4 - B8, B4 - B25, B4 - B27, and B4 - B28 deviated by more than 4.3~kJ~mol$^{-1}$ from experiment. Here, the highest absolute deviation was observed for molecule pair B1 - B4 with 4.9~kJ~mol$^{-1}$.
The Spearman correlation coefficient of the absolute deviations from experiment for the two simulation methods was relatively high at 0.88, indicating that some of the deviations might be related to shortcomings in the force field or in the experimental determination.

\begin{figure}[h]
    \centering
        \includegraphics[width=0.45\textwidth]{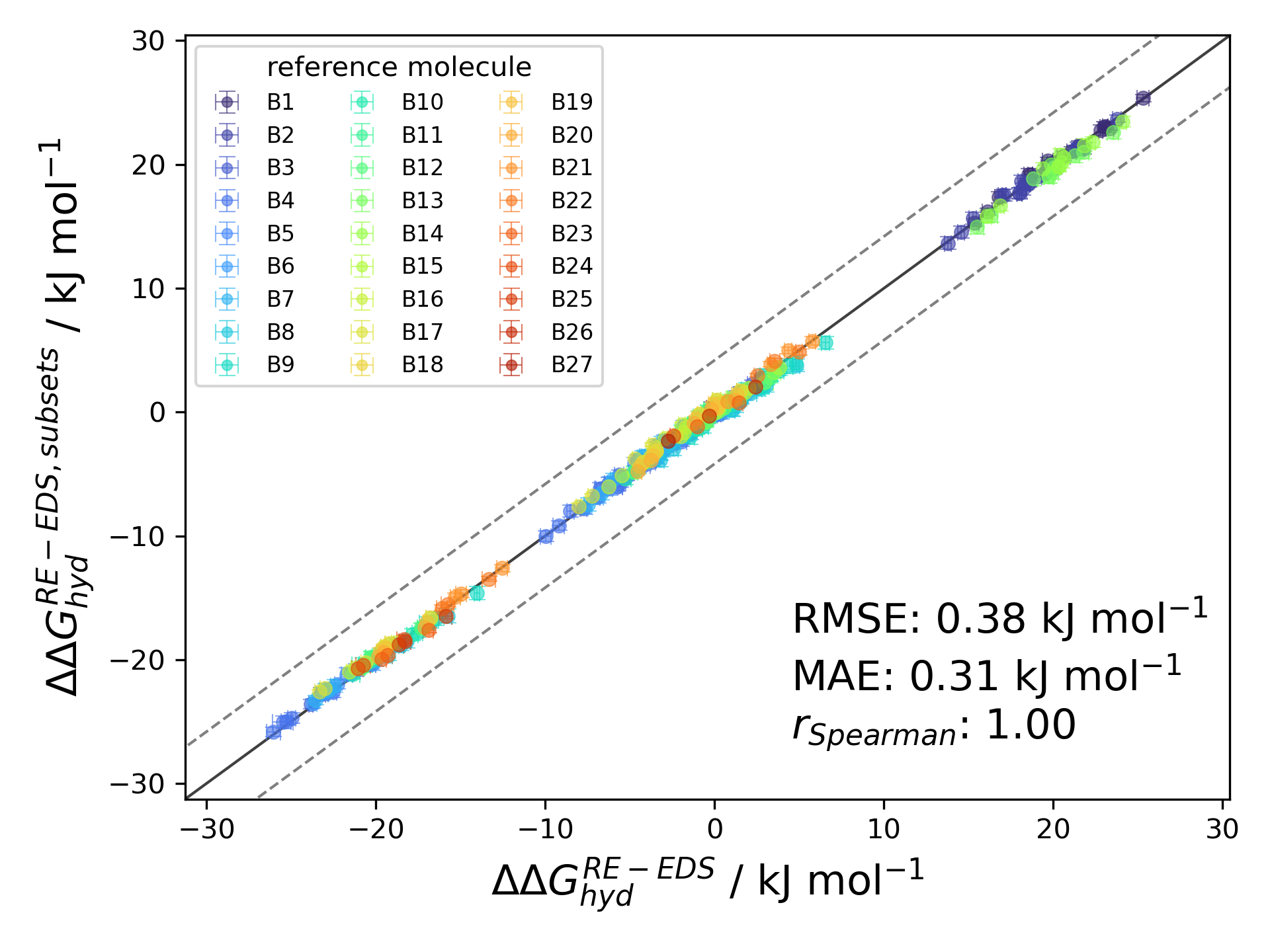}
        \includegraphics[width=0.45\textwidth]{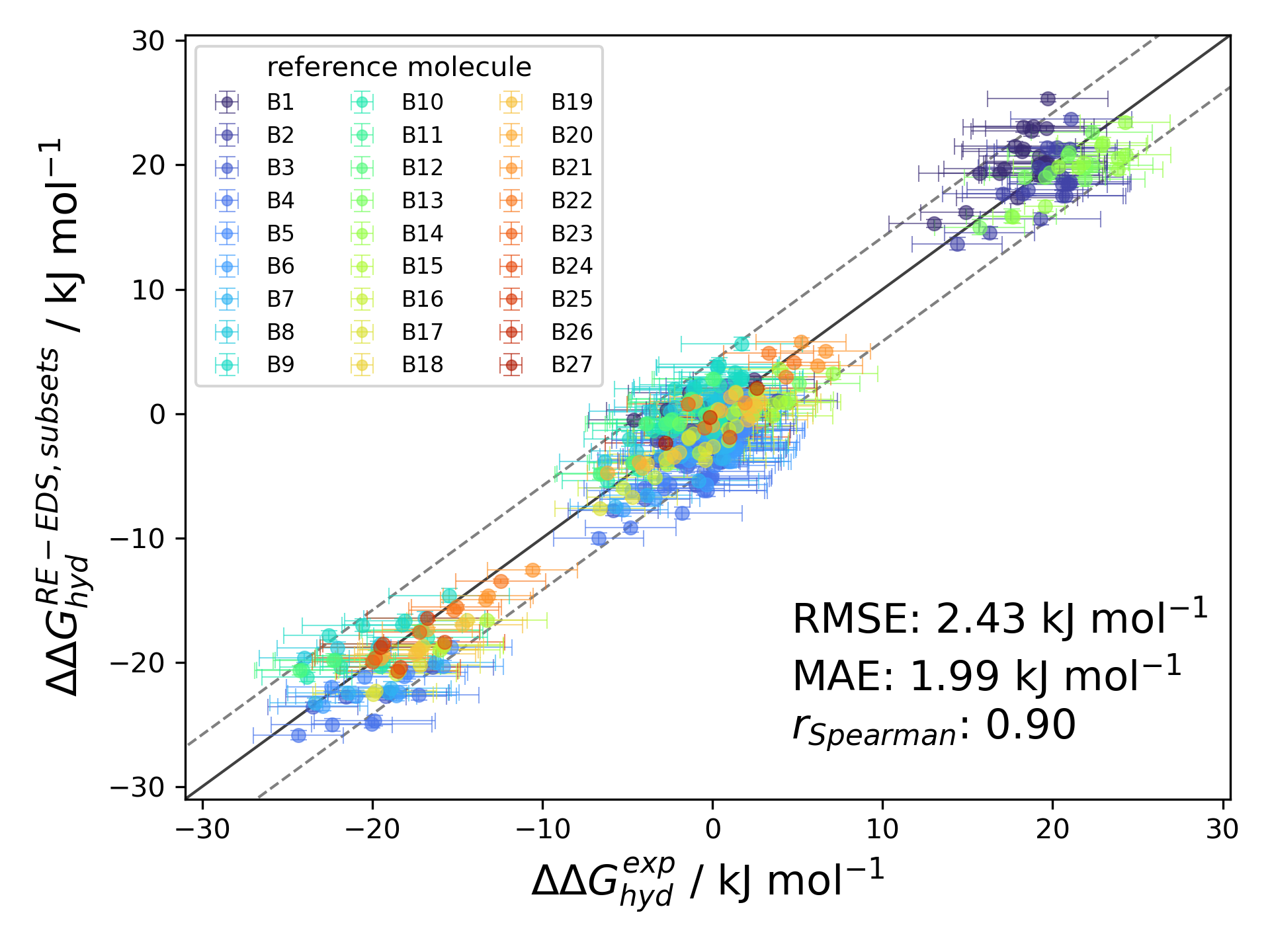}
    \caption{Comparison of the relative hydration free energies with RE-EDS of subsets Ba and Bb ($\Delta\Delta G_\text{hyd}^\text{RE-EDS,subsets}$) compared to the full set B ($\Delta\Delta G_\text{hyd}^\text{RE-EDS}$) (left), and compared to experiment ($\Delta\Delta G_\text{hyd}^\text{exp}$) (right) as reported by the FreeSolv\cite{Mobley2014,Duarte2017} database. The gray diagonal lines correspond to perfect alignment within $\pm$~4.184~kJ~mol$^{-1}$ ($\pm$~1~kcal~mol$^{-1}$). The $\Delta\Delta G_\text{hyd}^{ji}$ values are colored according to end-state $i$ (i.e., the ``reference molecule'' for the calculation). The RE-EDS results were averaged over five independent production runs in vacuum/water and the errors of the $\Delta G$ values correspond to the standard deviation over the five repeats. The error estimate of the $\Delta \Delta G$ values was calculated \textit{via} Gaussian error propagation. The numerical values are provided in Table S4 in the Supporting Information. A plot of the deviations from experiment for the different methods is shown in Figure S12 in the Supporting Information. All pairwise comparisons between the different simulation methods and the experimental results are provided in Figure S9.} 
    \label{fig: ddg_setB_subsets}
\end{figure}

To investigate the efficiency and accuracy of RE-EDS free-energy calculations for a larger number of molecules, the RE-EDS results of set B were compared to the ones obtained from RE-EDS simulations of subsets Ba and Bb (Figure \ref{fig: ddg_setB_subsets} and Table \ref{tab: summary_setB}). With an RMSE against the experimental values\cite{Mobley2014,Duarte2017} of 2.4~kJ~mol$^{-1}$, a MAE of 2.0~kJ~mol$^{-1}$, and a Spearman correlation coefficient of 0.90, the combined results from the two separate RE-EDS pipelines (i.e., Ba and Bb) were marginally more accurate. 
The agreement with the MBAR results\cite{Mobley2014,Duarte2017} was slightly higher with an RMSE of 0.8~kJ~mol$^{-1}$ and a MAE of 0.5~kJ~mol$^{-1}$ (Figure S9 in the Supporting Information). 
The RMSE between the RE-EDS results for set B versus subsets Ba and Bb was 0.4~kJ~mol$^{-1}$ with perfect correlation ($r_\text{Spearman}=1.00$). The slightly higher agreement of the results obtained from the two subsets with the experimental and MBAR results indicates that there is a small ``diffusion effect'' for this system when more molecules are added to the simulation. As more end-states are added to a system, the number of frames where an end-state contributes maximally to the reference state decreases. Additionally, more $s$-values are required to obtain frequent round-trips and more energy offset rebalancing iterations are needed to obtain approximately equal sampling of all the end-states.

Also here, the relatively small simulation time required for the RE-EDS calculations can be highlighted. The total simulation time for the RE-EDS simulations of set B was about 661~ns, compared to 2884~ns for MBAR\cite{Mobley2014,Duarte2017}. If pairwise TI simulations in vacuum/water had been used as for set A, the total simulation time would have been between 7128~ns (minimal 27 pairs = $N$-1) and 99627~ns (all pairs). The total simulation time for the RE-EDS pipelines of subsets Ba and Bb combined was about 643~ns. This is slightly shorter than the simulation length for the full set B, mainly due to the fact that for subset Bb in water, only four instead of eight replicas were added during the $s$-optimization, and one less rebalancing step was required for both subsets in vacuum and for Bb in water. Plots of the convergence of the free-energy calculations with RE-EDS can be found in Figures S10 and S11 in the Supporting Information. 

\begin{table}[h]
\caption{Set B: Overview of statistical metrics (RMSE, MAE, and Spearman correlation coefficients) with respect to the experimental results, and total simulation time for the different free-energy methods. The RE-EDS and TI results were averaged over five independent production runs in vacuum/water and the errors of the $\Delta G$ values correspond to the standard deviation over the five repeats. The error estimate of the $\Delta \Delta G$ values was calculated \textit{via} Gaussian error propagation.
For RE-EDS, both the results for the full set B and for the combined subsets Ba and Bb are reported.
The uncertainties of the RMSE and MAE values were estimated from the distribution of RMSE and MAE when a random selection of up to 26 molecules was removed from the calculations (5000 repetitions).
The accumulated simulation time is split into preparation (pre-processing, equilibration) and production time. The complete table can be found in Table S4 in the Supporting Information.} 

\label{tab: summary_setB}
\centering
\begin{tabular}{ | l | c | c | c | }
    \hline
    \multicolumn{1}{|c|}{} \rule{0pt}{12pt}& $\Delta\Delta G_\text{hyd}^\text{MBAR}$\cite{Mobley2014,Duarte2017} & $\Delta\Delta G_\text{hyd}^\text{RE-EDS}$ & $\Delta\Delta G_\text{hyd}^\text{RE-EDS,subsets}$\\ 
 \hline \hline
        {RMSE [kJ~mol$^{-1}$]} & $2.0~\pm~0.2$ & $2.6~\pm~0.3$ & $2.4~\pm~0.3$ \\
        {MAE [kJ~mol$^{-1}$]} & $1.6~\pm~0.2$ & $2.2~\pm~0.2$ & $2.0~\pm~0.2$ \\
        \hline
        {$r_{\text{Spearman}}$} & 0.92 & 0.89 & 0.90 \\
        \hline
        {$t_\text{preparation}$ [ns]} & 84~ns &  549~ns & 514~ns \\
        {$t_\text{production}$ [ns]} & 2800~ns  & 112~ns & 129~ns  \\
    \hline
\end{tabular}
\end{table}

%%%%%%%%%%% CONCLUSIONS %%%%%%%%%%%%%%%%
\section{Conclusion and Outlook}
In this work, the GROMOS++ program \textit{amber2gromos} was introduced to convert topologies from the AMBER to the GROMOS file format. An overview of the differences between AMBER and GROMOS force fields was presented together with a description of the conversion of the AMBER topology parameters to GROMOS topology parameters, and the necessary slight modification to the GROMOS source code. A workflow was outlined to prepare topology, coordinate, and distance restraint input files for RE-EDS free-energy calculations with GAFF parameters in GROMOS. The extension of this workflow to the OpenFF family of force fields is straightforward.

Two sets of benzene derivatives were selected from the FreeSolv database with six (set A) and 28 molecules (set B). Set A was used to validate the implementation of \textit{amber2gromos} and the related source-code changes to the GROMOS MD engine. The generated GROMOS topologies for the six benzene derivatives were compared to GROMACS topologies generated by ParmEd from the same AMBER topologies. Single-molecule simulations in vacuum performed in GROMOS and in GROMACS showed nearly identical energy and temperature distributions.

Finally, relative hydration free energies were calculated for both sets. For set A, both TI and RE-EDS simulations were carried out in vacuum/water to estimate the 15 pairwise free-energy differences. These results were compared to the relative hydration free energies obtained from experiment as well as the hydration free energies reported in the FreeSolv database (calculated with MBAR). Overall, an excellent agreement was observed between the different free-energy methods and with experiment. While all methods delivered highly accurate results, the RE-EDS calculations required the least amount of total simulation time.
The system size was increased to 28 molecules in set B to challenge the RE-EDS pipeline. Again, the results agreed well with the ones from MBAR and with the experimental values.

To test if it is more efficient to use a large set of molecules or two subsets with a shared molecule, set B was divided into two subsets Ba (molecules B1 - B14) and Bb (molecule B1 and molecules B15 - B28). 
While both the results and simulation time of the two RE-EDS approaches were almost identical, smaller subsets may offer some advantages in practice. RE-EDS simulations are in principle highly parallelizable, as large parts of both the replicas and the interactions within the replicas can be carried out independently with relatively infrequent communication. Nevertheless, as more molecules/replicas are added to the system, the wall-clock time of the simulations increases (more interactions to calculate, larger communication overhead, more replicas). Using two subsets decreased the elapsed real-time of the RE-EDS pipeline. Further research will be needed to determine optimal splits of datasets into subsets as well as the choice of the common molecule(s). The aim will be to find a balance between avoiding a diffusion effect from too many end-states in one system, and error propagation due to too small subsets or sub-optimal common molecule(s).

Overall, it has been shown that hydration free-energy calculations with RE-EDS and GAFF parameters executed in GROMOS accurately reproduce both experimental values and results obtained with different free-energy estimators and MD engines. While the molecules of the chosen datasets were relatively small and contained a well-defined common benzene core, previous studies successfully used RE-EDS to calculate binding and hydration free energies for molecule sets involving larger structural changes such as R-group modifications, ring opening/closing and ring size changes. 
In future work, GAFF parameterized topologies will be used to perform binding free-energy calculations with RE-EDS in GROMOS.

\section*{Data and Software Availability}
The input files for the RE-EDS simulations can be found at \url{https://github.com/rinikerlab/reeds/tree/main/examples/systems/benzenes_amber2gromos}. The GROMOS software package and the GROMOS++ package of programs can be downloaded for free at \url{http://gromos.net/}, AmberTools at \url{https://ambermd.org/AmberTools.php}, and GROMACS at \url{http://www.gromacs.org/}. The Python code for PyGromosTools and the RE-EDS pipeline are freely available at \url{https://github.com/rinikerlab}. \textit{amber2gromos} will be part of the next scheduled release of GROMOS.

\section*{Acknowledgements}
The authors gratefully acknowledge financial support by the Swiss National Science Foundation (Grant no. 200021-178762 and no. 200021-175944).

%%%REFERENCES%%%
\bibliographystyle{b2}
\bibliography{references}

\end{document}